# Domain Wall Magnetoresistance of Co Nanowires

R. F. Sabirianov[1,2], A. K. Solanki[2,3,4], J. D. Burton[2,3], S. S. Jaswal[2,3], and E. Y. Tsymbal[2,3]

[1]*Department of Physics, University of Nebraska, Omaha, NE 68182-0266*
[2]*Center for Materials Research and Analysis, University of Nebraska, Lincoln, NE 68588-0111*
[3]*Department of Physics and Astronomy, University of Nebraska, Lincoln, NE 68588-0111*
[4]*Central Computer Center, Malaviya National Institute of Technology, Jaipur 302017, India*

Using density functional theory implemented within a tight-binding linear muffin-tin orbital method we perform calculations of electronic, magnetic and transport properties of ferromagnetic free-standing *fcc* Co wires with diameters up to 1.5 nm. We show that finite-size effects play an important role in these nanowires resulting in oscillatory behavior of electronic charge and the magnetization as a function of the wire thickness, and a non-monotonic behavior of spin-dependent quantized conductance. We calculate the magnetoresistance (MR) of a domain wall (DW) modeled by a spin-spiral region of finite width sandwiched between two semi-infinite Co wire leads. We find that the DW MR decreases very rapidly, on the scale of a few interatomic layers, with the increasing DW width. The largest MR value of about 250% is predicted for an abrupt DW in the monatomic wire. We show that, for some energy values, the density of states and the conductance may be non-zero only in one spin channel, making the MR for the abrupt DW infinitely large. We also demonstrate that for the abrupt DW a large MR may occur due to the hybridization between two spin subbands across the DW interface. We do not find, however, such a behavior at the Fermi energy for the Co wires considered.

## I. INTRODUCTION

For a long time the electrical resistance of a magnetic domain wall (DW) in metallic ferromagnets has been attracting considerable interest (for a recent review see Ref.1). The origin of the DW resistance is attributed to the mixing of up- and down-spin electrons due to the mistracking of the electron's spin on passing through the DW.[2] The narrower DW width results in a larger angle between the magnetization directions of successive atomic layers thereby lowering the electron transmission and hence enhancing the resistance. In the ballistic regime, the change in resistance as a function of the DW width, $d_{DW}$, is determined by the electron Fermi wave length, $\lambda_F$. In bulk ferromagnets the DW width is entirely determined by the exchange and magnetic anisotropy energies and is typically $d_{DW} \sim 100$ nm, whereas $\lambda_F \sim 0.5$ nm. Hence, DWs do not affect appreciably the resistance of bulk ferromagnets because an electron can adiabatically follow the varying magnetization direction within the DW.

This behavior changes dramatically in magnetic nanostructures, where the reduced dimensions affect both the DW width and the mechanism of electron transport responsible for the DW resistance. For example, a very thin DW was predicted for atomic-size constrictions with the characteristic width of a few interatomic distances.[3] The enhanced DW resistance expected in magnetic nanostructures stimulated significant interest in the electronic transport through DWs due to new physics controlling the DW resistance and due to possible applications of the magnetoresistance (MR) associated with DWs in magnetoelectronic devices.

Recent advances in nanotechnology made it possible to measure a contribution to the resistance from a single DW.[4,5,6,7,8,9] Interestingly, the DW resistance turned out in some cases to be negative,[5,6] whereas in other cases to be positive.[4,7,8,9] Both results have found theoretical explanations.[10,11,12] Levy and Zhang[10] showed that diffuse scattering between electronic states of opposite spin orientation, which occurs in the process of electron transport across the rotating magnetization within a DW, leads to increased resistance. Tatara and Fukuyama[11] demonstrated that DWs can suppress weak localization due to the opening of additional conduction channels that results in a lower (negative) DW resistance. Van Gorkom *et al.*[12] found that the DW resistance could be either positive or negative, depending on the difference between the spin-dependent scattering rates due to the spatial variation of the magnetization value within the DW.

Constrained geometries of nanojunctions add new features to electronic transport. If the constriction size is less than or comparable to the mean free path, the conduction becomes ballistic rather than diffusive which is typical for bulk metals. When the constriction width is comparable to the electron Fermi wavelength, the electrical conductance is quantized. The quantized conductance was observed in metallic nanowires, where an atomic-size constriction is created by pulling apart two electrodes in contact (for a recent review see Ref.13). The conductance quantization can be explained within the Landauer formula,[14] and the adiabatic principle,[15] according to which the conductance is given by $\Gamma = Ne^2/h$, where $N$ is the number of open conducting channels i.e., the number of transverse modes at the Fermi energy. The conductance varies in discrete steps as the number of bands crossing the Fermi energy changes with the constriction width. For nonmagnetic nanowires the conductance is quantized in units of $2e^2/h$, where the factor 2 stands for spin degeneracy. If the constriction is made of a ferromagnetic metal, such as Ni, the exchange energy lifts the spin degeneracy and the conductance is quantized in units of $e^2/h$, provided the wire is uniformly magnetized. Such a phenomenon was observed in Ni break junctions,[16] Ni nanowires electrodeposited into pores of membranes,[17] Ni atomic-size contacts made by a scanning tunneling microscope,[18] and electrodeposited Ni nanocontacts grown by filling an opening in focused-ion-beam-milled nanowires.[19] Very recently Velev et al.[20] predicted an effect which they called ballistic anisotropic magnetoresistance (BAMR). Here the conductance of a narrow ferromagnetic wire changes in steps of $e^2/h$ when the magnetization is switched from along the wire to perpendicular to the wire.

The ballistic transport in ferromagnetic metal constrictions has recently received a great deal of attention due to unexpectedly large MR values obtained in experiments on Ni break junctions.[21] These results were attributed to a creation and annihilation of a constrained DW during a magnetic field sweep. Although the results of these experiments created significant controversy,[22] they stimulated a number of theoretical studies of spin-dependent transport in constrained geometries using free-electron models.[23,24,25] Imamura et al.[23] demonstrated that the interplay between quantized conductance and an atomic scale domain wall results in MR that oscillates with the cross section of the constriction and leads to enhanced MR values. The magnetoresistance fluctuations were also found by Tagirov et al.,[24] who used a quasiclassical approach to calculate the MR due to a constrained DW that was approximated by a step-like potential. Dugaev et al.[25] found an analytical solution for the MR of a narrow DW limiting their consideration of electronic transport to one quantum channel. Zhuravlev et al.[26] showed for atomic size constrictions that a closure of one spin conduction channel may result in very large magnetoresistance due to "half-metallic" behavior of the electrodes.

Although these free-electron theories provide a valuable insight into the DW resistance, they cannot be used for quantitative comparison with experiments due to the complex spin-polarized electronic structure of the ferromagnetic metals. It is well known that the band structures of transition metal ferromagnets are dominated by $d$ bands which cannot be properly described by a single parabolic band at the Fermi energy. Recent advances in band structure and electronic transport theory have made it possible to perform first-principles calculations of the DW MR. In particular, using the embedded Green-function technique based on a linearized augmented plane-wave method, Van Hoof et al.[27] carried out calculations of defect-free DWs in bulk Ni, Co, and Fe within the local spin-density approximation. They found a positive DW resistance with MR of about 0.1% for DW widths typical for bulk ferromagnets. Much higher MR values, i.e. 60-70%, were found by these authors for abrupt DWs. An even higher value of 250% was predicted for the abrupt DW in bulk fcc (001) Co by Kudrnovsky et al.,[28] who used a transmission matrix formulation of the conductance based on surface Green's functions within the tight-binding linear muffin-tin orbital method. They found that the DW MR drops down on a scale of a few interatomic distances as a function of the DW width. Yavorsky et al.[29] calculated the MR of a Fe superlattice with alternating regions of collinear and spiral-like magnetizations using a linearized Boltzmann equation within a state- and spin-independent relaxation time approximation.

All the above first-principles models of the DW MR have been applied to bulk ferromagnets and consequently have disregarded the lateral quantization of electronic waves which is decisive for electronic transport in nanowires and nanoconstrictions. Recently Velev and Butler[30] calculated the DW resistance in Ni, Co, and Fe nanocontacts using a semiempirical tight-binding approach. Bagrets et al.[31] and Solanki et al.[32] studied the magnetoresistance in metallic atomic-size constrictions using first-principles electronic structure methods.

In this paper, using fully self-consistent electronic structure obtained within density functional theory, we study electronic, magnetic and transport properties of ferromagnetic Co nanowires with diameters up to 1.5 nm. We show that finite-size effects play an important role resulting in (i) oscillatory behavior of the electronic charge and magnetic moments within the wires, (ii) a non-monotonic variation of the magnetization as a function of wire thickness, (iii) spin-dependent conductance quantization reflecting the electronic structure of the wires, and (iv) a non-monotonic change in the DW MR with increasing wire thickness. We demonstrate that, for some electron energy values, the conductance may display half-metallic behavior reflecting non-zero density of states only within one spin channel. Additionally, we show that large MR can be observed for the abrupt DW due to the hybridization between two spin subbands.

## II. METHOD OF CALCULATION

We consider free standing, translationally invariant nanowires of ferromagnetic fcc cobalt. The nanowires are built along the [001] direction (z axis) by periodic repetition of a supercell made up of two fcc (001) planes (except for the monatomic wire). We consider five nanowire configurations having different atomic arrangements: (i) monatomic, i.e. infinite 1D chain of atoms, (ii) 2×2, (iii) 5×4, (iv) 13×12, and (v) 25×24. To take advantage of the k-space representation within a first-principles calculation, we consider a periodic array of these wires separated by empty space as described below.

A monatomic Co wire is built assuming that it lies along the face diagonal of an *fcc* lattice. The resultant unit cell is a body-centered tetragonal unit cell with $a = a_{fcc}/\sqrt{2}$ and $c = a_{fcc}$, where $a_{fcc}$ = 6.703 a.u. is the lattice parameter of bulk *fcc* Co. The periodic array of monowires has a spacing of 3 unit cells between the wires to minimize the interactions between them.

The 2×2 wire is modeled by a super-cell of two *fcc* (001) layers. Each layer has 18 sites (large enough to separate it from the rest of the array) with only two sites in each layers occupied by Co atoms while the rest are kept empty. This forms a wire with a 4 atom square cross section. Similarly, a 5×4 wire (the cross-section of which is shown in Fig.1) is modeled by two *fcc* (001) layers. Each layer has 25 sites such that one layer has 5 Co atoms and the next has 4. The rest are empty spheres. In a similar way we build the 13×12 wire with 25 = 13 + 12 sites occupied by Co atoms and 24 empty spheres surrounding the cell. Our largest 25×24 wire has 98 sites with 49 Co atoms in two *fcc* layers. This wire has a square cross section of about 1.5 × 1.5 nm.

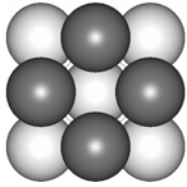

**FIG. 1.** Cross-section of the 5×4 wire representing a periodically-repeated super cell of two *fcc* (001) layers with 5 (white) and 4 (grey) Co atoms in each layer.

The spin-polarized electronic band structure of the Co nanowires is calculated self-consistently using density functional theory implemented in a tight-binding linear muffin-tin orbital (TB-LMTO) method within the atomic sphere approximation (ASA). For uniformly magnetized wires we calculate the electronic structure in *k*-space. In all our calculations we disregard the spin-orbit interaction and neglect any structural relaxation.

A DW is modeled by a spin-spiral region of finite width such that the angle between the magnetic moments of two successive atomic layers is constant, and the magnetic moments of individual atoms are collinear within each atomic layer. The DW is confined within the region between two Co semi-infinite leads having antiparallel magnetization orientations. In the presence of a DW the electronic structure and the conductance are calculated in real space. For the transport calculations, we use the self-consistent electronic potential obtained in each case to produce the Hamiltonian $H$ for each of the semi-infinite Co leads and the Hamiltonian $H_S$ for the scattering region containing the central sites with the DW and three layers from each lead. First we calculate the surface Green function for the left (L) and right (R) semi-infinite leads, $G_L$ and $G_R$, by solving the equations

$$G_L = \left(E - H - V_L G_L V_L^\dagger\right)^{-1}, \quad (1)$$

$$G_R = \left(E - H - V_R^\dagger G_R V_R\right)^{-1}, \quad (2)$$

where $E$ is the electron energy and $V_{L,R}$ describe the hopping to and from the barrier for the right ($R$) and left ($L$) lead. The Hamiltonian $H_S$ of the scattering region is built from the self-consistent potentials which must be transformed from their local spin quantization axis, defined by the direction of the magnetic moment, to the global *z*-axis. This involves the unitary transformation of the layer-dependent potential parameters $P_n$ as follows

$$P_n = U^\dagger(\theta_n, \varphi_n) P_0 U(\theta_n, \varphi_n). \quad (3)$$

Here the rotation matrices $U(\theta_n, \varphi_n)$ are:

$$U(\theta_n, \varphi_n) = \begin{pmatrix} e^{i\varphi_n/2}\cos\frac{\theta_n}{2} & e^{-i\varphi_n/2}\sin\frac{\theta_n}{2} \\ -e^{i\varphi_n/2}\sin\frac{\theta_n}{2} & e^{-i\varphi_n/2}\cos\frac{\theta_n}{2} \end{pmatrix}. \quad (4)$$

The Green function of the total system, i.e. the DW coupled to the leads, is given by

$$G = \left(E - H_S - \Sigma_L^\dagger - \Sigma_R\right)^{-1}, \quad (5)$$

where $\Sigma_L$ and $\Sigma_R$ are the self energies associated with the left and right leads respectively. The conductance $\Gamma$ is calculated using the Landauer-Büttiker formula[14,33]

$$\Gamma = \frac{e^2}{h} T, \quad (6)$$

where $T$ is the transmission coefficient summed up over all the incoming and outgoing electronic states of the left and right leads. At zero bias voltage and zero temperature the transmission coefficient can be found from the Green function $G(E_F)$ taken at the Fermi energy $E_F$:[34]

$$T = Tr\left[\left(\Sigma_L^\dagger - \Sigma_L\right) G(E_F) \left(\Sigma_R - \Sigma_R^\dagger\right) G^\dagger(E_F)\right]. \quad (7)$$

The self energies are expressed through the hopping integrals and the surface Green functions of the uncoupled electrodes, $G_L$ and $G_R$, as follows:

$$\Sigma_R = V_R G_R V_R^\dagger, \quad (8)$$

$$\Sigma_L = V_L^\dagger G_L V_L. \quad (9)$$

The conductance of a magnetically-saturated nanowire, $\Gamma_0$, is different from the conductance of the nanowire in the presence of the DW, $\Gamma_{DW}$. We define the DW MR value by the ratio

$$MR = \frac{\Gamma_0 - \Gamma_{DW}}{\Gamma_{DW}}. \quad (10)$$

In addition to the first-principle approach, we use a simple one-dimensional tight-binding (TB) model to provide a simple analysis of the DW MR. A single-band TB

Hamiltonian takes the form: $H = V - \Delta\sigma_z$, where $V$ is the hopping integral which is assumed to be non-zero only between nearest-neighbor atoms, $\Delta$ is the Stoner exchange splitting parameter and $\sigma_z$ is the Pauli matrix. The magnetization variation within the DW is obtained by the unitary transformation $\tilde{\sigma} = U^\dagger \sigma_z U$ which is performed on each site.

We use a one-band TB model to predict the upper limit for magnetoresistance. In this model the bandwidth is determined by the hopping integral $V$. If this parameter is small, the neighbors interact weakly and states are, to a large degree, localized on each site. The exchange parameter $\Delta$ controls the splitting of the band between majority- and minority-spin states. When $\Delta$ is larger than $V$ and the band is half-filled, the Fermi energy lies within the majority spin band and the minority spin band gap. This case corresponds to a half-metallic magnet which is expected to have the largest DW MR value.

A two-band model can be built in a similar fashion. The difference is that two bands are allowed with different hopping parameters $V$. One band is made wide to simulate s-like states and another one is narrow to mimic d-like states. This model is used below to explain the appearance of gaps in the spectrum of electronic states obtained for Co nanowires from TB-LMTO calculations.

**Table 1.** Electronic, magnetic and transport properties of Co nanowires: charge transfer, $\Delta q$, magnetic moment per atom ($m$), number of bands ($N$) crossing the Fermi energy for majority (maj) and minority (min) spin electrons, ballistic conductance per unit area ($\Gamma/A$), and MR for an abrupt DW. ($A$ for the monatomic wire is chosen as ¼ the area of the 2×2 wire.) $<m>$ denotes an average magnetic moment per atom. For bulk Co ballistic conductance values are taken from ref. 41 and the abrupt DW MR value from ref. 28. Charge neutrality is maintained when the charge transfer to the empty spheres is taken into account. $r_0$ is the radius (in units of $a_{fcc}/2$) from the axis of the wire for each atom type.

| Type of wire | | | $\Delta q$ ($e$) | $m$ ($\mu_B$) | N | | $\Gamma_0/A$ ($10^{15}$ $\Omega^{-1}$m$^{-2}$) | | $\Gamma_{DW}/A$ ($10^{15}$ $\Omega^{-1}$m$^{-2}$) | MR (%) |
|---|---|---|---|---|---|---|---|---|---|---|
| | | | | | min | maj | min | maj | | |
| **Monatomic** | | | 0.65 | 2.31 | 6 | 1 | 1.42 | 0.24 | 0.47 | 253 |
| **2 x 2** | | | 0.32 | 1.84 | 3 | 3 | 1.78 | 1.79 | 3.36 | 6 |
| **5 x 4** | | | | | 6 | 5 | 0.89 | 0.74 | 1.48 | 10 |
| | Type | $r_0$ | -0.43 | 1.43 | | | | | | |
| Layer 1 | Atom 1 | 0.00 | 0.42 | 1.78 | | | | | | |
| | Atom 2 | 1.00 | 0.17 | 1.75 | | | | | | |
| Layer 2 | Atom 3 | 1.00 | | $<m> = 1.72$ | | | | | | |
| **13 x 12** | | | | | 8 | 7 | 0.30 | 0.26 | 0.49 | 14 |
| Layer 1 | Atom 1 | 0.00 | 0.06 | 1.66 | | | | | | |
| | Atom 2 | 1.00 | -0.21 | 1.48 | | | | | | |
| | Atom 3 | 1.41 | 0.26 | 1.86 | | | | | | |
| | Atom 4 | 2.00 | 0.42 | 1.77 | | | | | | |
| Layer 2 | Atom 5 | 1.00 | -0.09 | 1.71 | | | | | | |
| | Atom 6 | 1.73 | 0.21 | 1.79 | | | | | | |
| | | | | $<m> = 1.73$ | | | | | | |
| **25 x 24** | | | | | 19 | 10 | 0.31 | 0.17 | 0.25 | 92 |
| Layer 1 | Atom 1 | 0.00 | 0.00 | 1.72 | | | | | | |
| | Atom 2 | 1.00 | 0.02 | 1.77 | | | | | | |
| | Atom 3 | 1.41 | -0.05 | 1.73 | | | | | | |
| | Atom 4 | 2.00 | -0.12 | 1.69 | | | | | | |
| | Atom 5 | 2.24 | 0.16 | 1.83 | | | | | | |
| | Atom 6 | 2.83 | 0.54 | 1.78 | | | | | | |
| Layer 2 | Atom 7 | 1.00 | 0.01 | 1.66 | | | | | | |
| | Atom 8 | 1.73 | -0.09 | 1.70 | | | | | | |
| | Atom 9 | 2.24 | -0.04 | 1.67 | | | | | | |
| | Atom 10 | 2.65 | 0.41 | 1.85 | | | | | | |
| | | | | $<m> = 1.75$ | | | | | | |
| **Bulk** | | | 0 | 1.67 | | | 1.12 | 0.47 | 0.45 | 253 |

## III. ELECTRONIC AND MAGNETIC STRUCTURE

The electronic structure of Co nanowires is quite different compared to that of bulk Co due to the large number of atoms at the surface. The reduced coordination for these atoms leads to sizable charge transfers and enhanced magnetic moments for these atoms. Table 1 shows the electronic and magnetic structure results which include charge transfers, $\Delta q$, and magnetic moments, $m$, for all the considered geometries of the nanowires and for bulk $fcc$ Co. An increase in electron occupation, relative to the atomic state, is denoted by $\Delta q < 0$, and $\Delta q > 0$ implies the atom has lost electrons.

As is seen from Table 1, a monatomic Co wire shows appreciably enhanced magnetic moment per atom, $m = 2.31\mu_B$, compared to the bulk value of $1.67\mu_B$. This result is in agreement with the experimental and other theoretical findings.[35,36] For the 2×2 wire configuration all four constituent atoms in the two planes of the supercell are of the same type due to their identical environment. From Table 1 we see again a sizable enhancement of the magnetic moment, $m = 1.84\mu_B$, due to an atomic-like environment with very few Co neighbor atoms to hybridize with.

For the 5×4 wire configuration, Co atoms can be classified into three different types within the two layers of the supercell according to tetragonal symmetry. For the 13×12 and 25×24 wire configurations, Co atoms can be classified into six and ten types, with the two $fcc$ (001) planes having 4 and 2 types for the 13×12 wire and 6 and 4 types for the 25×24 wire, respectively. For these wires the outermost atoms with lowest coordination have a substantial charge transfer towards the first nearest neighbor inside the wire. For example, the 4 atoms of type 6 (surface corner atoms) in the 25×24 wire lose electrons with $\Delta q = 0.54e$ in the atomic sphere while the 4 atoms of type 4 acquire electrons $\Delta q = -0.12e$. This implies an oscillatory behavior in the charge transfer when moving from surface atoms to core atoms.

The charge oscillations correlate strongly with the magnetic moment variations: the atoms which gain electrons have lower magnetic moments while the atoms which lose electrons have larger magnetic moments compared to the average moment of the wire. In particular, atoms located close to the center of the wire have local magnetic moments close to the bulk value. Nearly all atoms that lose electrons have moments above the average moment of the wire. Corner atoms have magnetic moments above $1.8\mu_B$.

The direct correlation between $\Delta q$ and $m$ can be explained by the fact that the minority-spin density of states (DOS) at the Fermi energy is much higher than the majority-spin DOS which is a consequence of the partially filled $d$ band for the minority-spin electrons. This is evident from Fig. 2a which shows the DOS for a monatomic Co wire. Gaining electrons by an atom implies filling the minority $d$ band that reduces the magnetic moment of this atom, whereas losing electrons implies depopulation of the minority $d$ band that enhances the magnetic moment. Similar oscillatory behavior of magnetic moments is known from the studies of electronic properties of ferromagnetic metal surfaces.[37]

Interestingly, for the 25×24 wire, with an approximate side length of 1.5 nm, the average magnetic moment, $<m>$, is larger than that for the 5×4 and 13×2 wires ($<m> = 1.75\mu_B$ versus $<m> = 1.72\mu_B$ and $<m> = 1.73\mu_B$, respectively). This is because the 25×24 configuration has the larger number of atoms which lose electrons compared to the other two geometries.

We find that the magnetization varies in an oscillatory fashion with increasing wire cross section. This is similar to the behavior observed for free clusters.[38] There are two reasons for this oscillation to occur. The first reason is the discontinuous variation of the number of core and surface atoms with the filling of the successive atomic shells as the wire thickness increases. The variation of the Co moments in the outermost atomic shell is due to the changing Co coordination number as determined by symmetry. The second reason is the charge and spin density oscillations across the wire. The charge density creates a standing wave due to the confinement effect similar to that predicted within the jellium model.[39] The charge oscillations in nanowires are more pronounced than the respective charge oscillations near the surface of a semi-infinite metal. The charge density oscillations lead to spin-density oscillations in the manner described above. A change in the cross-sectional area of the wire modifies the pattern of these oscillations. As a result the magnetization of the wire changes in an oscillatory fashion. We expect that this oscillatory trend in the magnetization will continue with increasing thickness of wires and stabilize eventually at the bulk magnetic moments.

## IV. CONDUCTANCE AND MAGNETORESISTANCE

Due to the periodicity of the wires along the $z$ direction, the ballistic conductance of a uniformly magnetized wire is solely determined by the number of bands, $N$, crossing the Fermi energy ($E_F$) along the wire direction. This is the consequence of the transmission coefficient being equal to unity for each conduction channel due to no reflection or mixing of spin channels of incoming electronic waves. We calculated band dispersions along the direction of the wire and found the number of bands crossing the Fermi energy. The results are shown in columns 4 and 5 of Table 1 for minority- and majority-spin electrons respectively. For a monatomic wire there is a large spin asymmetry in the number of bands crossing the Fermi energy: six majority-spin bands cross $E_F$ compared to only one minority-spin band. This result is similar to that obtained by Smogunov $et.$ $al.$[40] This asymmetry disappears for the 2×2 wire, for which there are three bands crossing $E_F$ in both spin channels. The 5×4, 13×12, and 25×24 wires display some spin asymmetries in $N$ which vary with the cross section of the wire.

The spin-dependent ballistic conductance is given by $\Gamma = Ne^2/h$. We calculated the ballistic conductance per unit area by dividing Γ with the cross sectional area of the nanowires, which allows comparison with the values of the conductance obtained for the wires to the value known for bulk Co.[41] As is evident from Table 1 (see columns 6 and 7), the ballistic conductance per unit area varies appreciably with the nanowire thickness displaying strong non-monotonic behavior. This variation reflects features of the electronic band structure of the nanowires. With increasing thickness of the wires one expects that the spin conductance will eventually saturate at the bulk value given in Table 1. In this limiting case the ballistic conductance is simply proportional to the cross section of the wire. However, for the wires in the *nm*-thickness range, we find a significant departure of the conductance values from those in the bulk. Even for the 25×24 wire we find that the conductance differs by a factor of more than three from the bulk value. This fact indicates the importance of the adequate description of the band structure for the prediction of electronic transport properties of wires in a nanometer range of thickness.

We note that for all cases (except for the 2×2 wire) minority-spin electrons have a larger $N$ compared to majority-spin electrons. This reflects the presence of the $d$ bands at the Fermi energy in the minority-spin channel (see Fig. 2a) which makes the DOS and the ballistic conductance of this spin channel higher. This is different from the diffusive regime in which majority-spin electrons have much higher conductivity due to the dispersive $s$-$p$ bands crossing the Fermi energy.[42,43]

The conductance variation as a function of energy reflects features of the electronic band structure of the wires. Fig. 2b shows the conductance Γ for majority- and minority-spin electrons for a monatomic uniformly magnetized Co wire. As expected, Γ is quantized in units of $e^2/h$, reflecting the changing number of open conducting channels, i.e. the number of bands crossing the appropriate energy. This picture correlates with the DOS shown in Fig.2a: if the energy lies within the $d$ band having much larger DOS, the conductance is higher, whereas if the energy lies within the s band the conductance is lower.

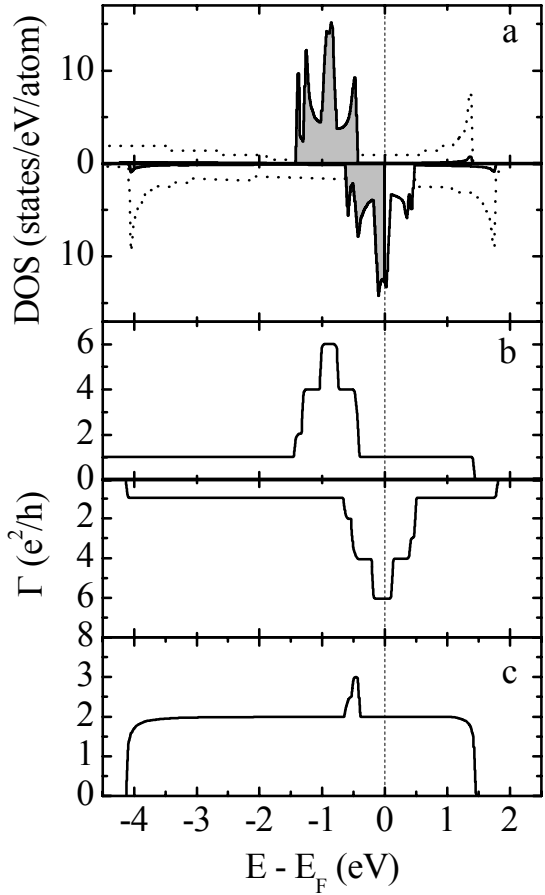

**FIG. 2.** a) Density of states for monatomic Co wire for majority- (the top panel) and minority- (the bottom panel) spin electrons as a function of energy. The dotted curve is the s-p partial DOS scaled by a factor of 10 to make it visible. b) Conductance of a ferromagnetic wire as a function of energy for majority- and minority- spin channels. c) Conductance of the abrupt DW as a function of energy. The Fermi energy is denoted by the dashed vertical line.

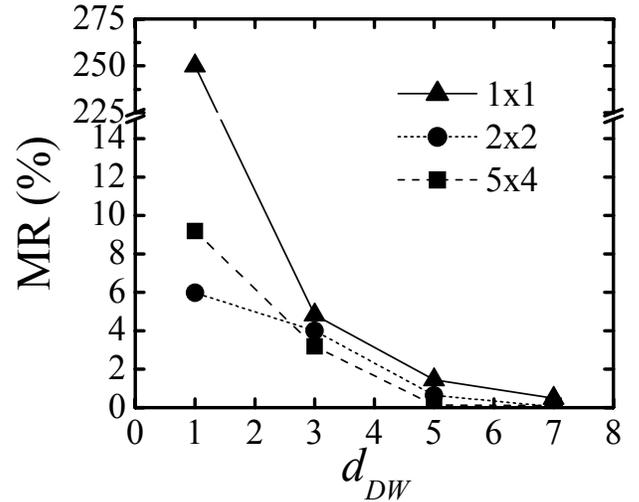

**FIG. 3.** Domain-wall magnetoresistance as function of the domain-wall width, $d_{DW}$ in units of the interlayer separation, for a monatomic (triangles), 2x2 (squares), and 5x4 (circles) wires.

It may happen that, for certain energies, there is a gap in one of the spin DOS making its spin-conductance equal to zero. This indeed occurs for the monoatomic Co wire for energies lying just above the top of the majority-spin band and just below the minority-spin band (see the top and bottom panels in Fig. 2b). If these energies were the Fermi energy, the ferromagnetic metal would behave as a half metal, i.e. material for which only one spin band is occupied, resulting in a 100% spin polarization.[44] In the case of a half-metal the electronic conduction through an abrupt domain wall is blocked by the spin conservation rule.[26] Indeed, if the magnetizations of two adjacent domains are antiparallel the spin channel that is open in the left domain is closed in the right domain and vice versa. This makes the conductance between the antiparallel-aligned leads with the abrupt magnetization change equal to zero. This is opposite to the case of the parallel-aligned leads for which one spin channel is open and the conductance is not equal to zero. Our calculations do not predict, however, the true half-metallic behavior for the Co wires considered. At least one band is always present at the Fermi energy in each spin channel, the spin conductance gap opening being possible only for energies different from the Fermi energy.

As was shown previously for bulk Co,[28] the DW MR drops down with increasing DW width, $d_{DW}$, on a scale of a few interatomic distances. We find a similar behavior for Co wires, although both the MR values and the conductance variation as a function of $d_{DW}$ vary significantly depending on the cross section of the nanowires. Fig. 3 shows results for the monatomic, 2x2, and 5x4 wires. We see that despite the sizable difference in the absolute MR values for the three wires, in all the cases the MR drops on a length scale of 2-4 interlayer distances.

The fast decrease of the DW MR as a function of the DW width can be qualitatively understood using a simple one-dimensional single-band tight-binding model described in Sec. II. We find that within this model the DW MR becomes very small for the DW width more than 3-5 atomic layers. This is the case even if one spin channel does not have any states at the Fermi energy, i.e. the ferromagnet is a half-metal.

This result can be understood using an analogy with an optical polarizer. If two ideal polarizers are at 90° to each other, there is no light coming through. But if another polarizer at 45° is inserted between them, the light can go through with 1/4 intensity of incident light. Inserting a few polarizers with a gradual change in angle will result in almost no loss in the light transmission (only the polarization direction will change).

For a half-metallic ferromagnet, a single-band tight-binding model gives the largest MR value in a narrow band limit. In this case the transmission coefficient $T$ across an abrupt DW between two leads with the magnetization direction rotated by angle $\theta$ is given by

$$T = \frac{4\cos^2\left(\frac{\theta}{2}\right)}{\left(1+\cos^2\left(\frac{\theta}{2}\right)\right)^2}. \qquad (11)$$

Fig. 4 shows the transmission coefficient and the MR for this interface. This behavior is reminiscent of the Malus' law in optics,[45] but the angle is divided by a factor of two and there is an additional angle-dependent denominator which comes from the propagator in the Landauer-Büttiker formalism. If we consider the DW as a collection of these abrupt interfaces with relative angle $\pi/n$, where $n$ is an integer and represent the number of atomic layers in the DW, then the transmission coefficient, $T$, approaches unity very fast with increasing $n$. Note that the transmission coefficient is almost equal to unity within the interval of angles from 0 up to about $\pi/2$, and then the $T$ drops abruptly to zero (see Fig.4). It means that the MR is quite small when the relative angle between the directions of the local moments in the consecutive monolayers of the wire is smaller than $\pi/2$. This corresponds to 3-5 monolayers. Note, that this is the upper limit for MR. Realistic bands with finite bandwidth would give smaller MR values. Therefore, for a large MR the DW should be abrupt representing a sharp flip in the magnetization direction.

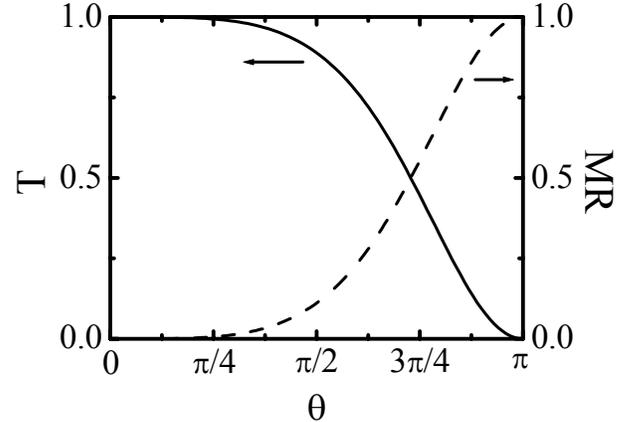

**FIG. 4.** Transmission coefficient (solid line) and magnetoresistance (dashed line) of an abrupt DW between two half-metallic electrodes with the magnetization direction rotated by angle $\theta$ as predicted by a one-dimensional single-band tight-binding model in a narrow band limit. Note that the MR is defined here by $MR = (\Gamma_0 - \Gamma_{DW})/\Gamma_0$ so that the maximum MR value is equal to unity.

The electronic structure of Co nanowires which strongly depends on the wire cross section has a dramatic effect on the DW MR. As is evident from Fig. 3, the MR values vary strongly for Co nanowires of different cross section. In particular in the case of the abrupt DW, in which the magnetic moment orientation changes from parallel to antiparallel within 1ML of Co, the largest MR value of 250% is predicted for a monatomic wire, whereas it is much smaller for 2x2 and 5x4 wires (6% and 10% respectively). Interestingly, the MR shows a very non-monotonic behavior with increasing cross sectional area of the wires. As is seen from Table 1, the MR value obtained for an abrupt DW is higher for 13x12 and 25x24 wires (15% and 90% respectively) than for 2x2 and 5x4 wires. This variation in the MR values reflects changes in the electronic structure of the Co wires. Table 1 indicates that there is a strong correlation between the asymmetry in the number of bands, $N$, crossing the Fermi energy for majority and minority spin electrons for uniformly-magnetized wires and the MR values. For example, the highest MR values obtained for monatomic and 25x24 wires is the consequence of the largest ratios of open spin channels for these wires. Surprisingly, the predicted value of about 250% obtained for the abrupt DW MR in bulk *fcc* (001) Co [28] is as large as the value we predict for a monatomic Co wire. We note, that this value is reduced to 67% for abrupt DW MR in bulk *fcc* (111) Co.[27]

Half-metallic behavior is not the only case when large MR can be observed. As is evident from Figs. 5a and 5b, for the 2x2 wire there are no gaps in the minority- or the majority-spin bands near the Fermi energy. However, Fig. 5c demonstrates that for the abrupt DW the conductance is strongly suppressed in the region about 0.3eV above the Fermi energy (this is indicated in Fig. 5c by the arrow). It appears that in this case the electronic hybridization in the antiparallel alignment leads to the "pseudogap" in the density of states. The mechanism which causes the suppression of the conductance in the antiparallel configuration in systems that are metallic in the ferromagnetic configuration is different from the "half-metallic" mechanism discussed above.

uniformly-magnetized wire there is no band gap in the density of states. This leads to the conductance of the majority and minority spin electrons showing no reduction within the band region (Fig. 6b). For the wire with the abrupt DW, however, there is a coupling between states in the one spin channel and states in the other spin channel across the DW. In this case if there are two states with similar on-site energy, they hybridize in such a way that the bonding and anti-bonding levels appear with the splitting of the order of the hybridization parameter. This causes the band to split into two subbands with the gap between them. This creates a pseudogap in the conductance across the abrupt DW at these energies (see Fig. 6c). This statement remains valid also if there are extended (s-like) states in both spin channels in the ferromagnetic state. Thus, for the abrupt DW a large magnetoresistance can occur due to the hybridization between the two spin bands across the DW interface.

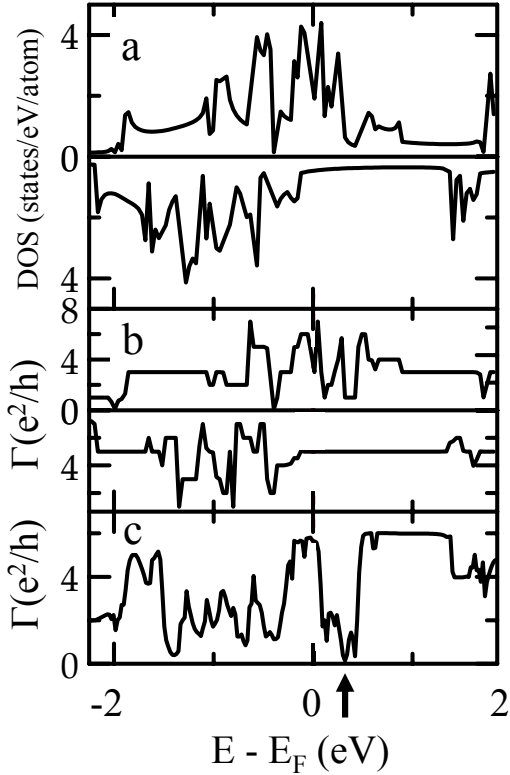

**FIG. 5.** a) Density of states for 2x2 Co wire for majority- (the top panel) and minority- (the bottom panel) spin electrons as a function of energy. b) Conductance of a ferromagnetic wire as a function of energy for majority- and minority- spin channels. c) Conductance for the abrupt DW configuration as a function of energy. The vertical arrow shows the energy at which the conductance through the abrupt DW is strongly suppressed.

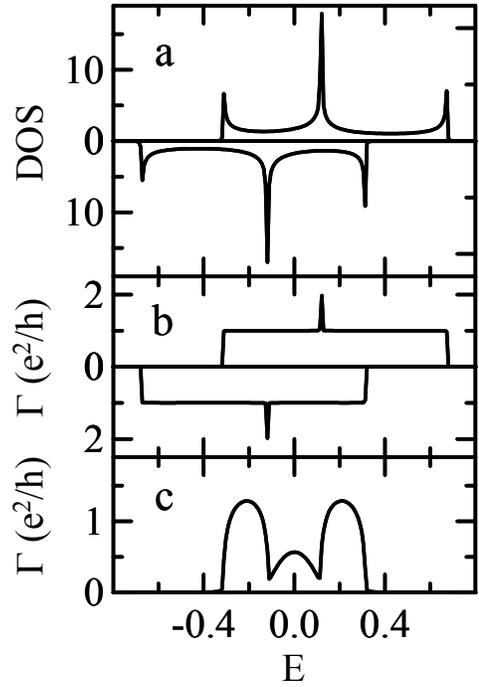

**FIG. 6.** Results of a two-band tight-binding model: a) Density of states for minority- (the top panel) and majority- (the bottom panel) spin electrons as a function of energy. b) Conductance for minority - and majority -spin channels as a function of energy. c) Conductance for the abrupt DW as a function of energy.

## V. CONCLUSIONS

Using density functional theory implemented within a tight-binding linear muffin-tin orbital method we have performed calculations of the electronic, magnetic and transport properties of ferromagnetic free-standing *fcc* Co wires oriented in the [001] direction with diameters up to 1.5 nm. We found that there is a substantial redistribution of charge, creating a charge density standing wave across the wire. These charge oscillations correlate strongly with the

This origin of this behavior can be understood within a simple tight-binding model with two bands of a different bandwidth. In order to mimic the d-metal we choose one band to be wide (with large hopping integrals), and one to be narrow. In the ferromagnetic state the up- and down-spin bands are exchange split. As is seen from Fig.6a, for a magnetic moment variations: the atoms which gain electrons have lower magnetic moments while the atoms which lose electrons have larger magnetic moments compared to the bulk value. The magnetization of the Co wires oscillates with increasing wire thickness similar to that observed for free ferromagnetic nanoparticles.

The ballistic conductance of the nanowires was calculated using Landauer-Büttiker formalism. We found that the conductance of uniformly-magnetized wires per unit cross sectional area varies in a non-monotonic fashion reflecting features of the electronic band structure and differs from the ballistic conductance for bulk *fcc* Co. We modeled a domain wall (DW) by a spin-spiral region of finite width placed between antiparallel-aligned Co leads and calculated the DW magnetoresistance (MR). We found that the predicted DW MR varies non-linearly as a function of the wire thickness and decreases very rapidly, on a scale of a few monolayers of *fcc* (001) Co, with increasing DW width. The latter behavior is explained in terms of the angular dependence of the conductance through an abrupt interface between two semi-infinite leads with magnetization directions rotated by a finite angle. The largest MR value of about 250% is predicted for an abrupt DW in a monatomic Co wire. The variation of the DW MR as a function of electron energy is very sensitive to the electronic structure of the wire. We found that for some energy values the conductance displays half-metallic behavior making the MR of an abrupt DW for these energies infinitely large. Also we showed that for the abrupt DW a large MR can occur due to the hybridization between two spin subbands across the DW interface. We did not find, however, such a behavior at the Fermi energy for the Co wires considered.

## ACKNOWLEDGEMENTS

This work is supported by National Science Foundation (grants DMR-0203359 and MRSEC: DMR-0213808), the Office of Naval Research (grant N00140210610), and the Nebraska Research Initiative. The calculations were performed using the Research Computing Facility of the University of Nebraska-Lincoln.[1] A. D. Kent, J. Yu, U. Rüdiger, and S. S. P. Parkin, J. Phys.: Cond. Matt. **13**, R461 (2001).
[2] G. G. Cabrera and L. M. Falicov, Phys. Stat. Sol. (b) **61**, 539 (1974).
[3] P. Bruno, Phys. Rev. Lett. **83**, 2425 (1999).
[4] J. F. Gregg, W. Allen, K. Ounadjela, M. Viret, M. Hehn, S. M.Thompson, and J. M.D. Coey, Phys. Rev. Lett. **77**, 1580 (1996).
[5] U. Ruediger, J. Yu, S. Zhang, A. D. Kent, and S. S. P. Parkin, Phys. Rev. Lett. **80**, 5639 (1998).
[6] T. Taniyama, I. Nakatani, T. Namikawa, and Y. Yamazaki, Phys. Rev. Lett. **82**, 2780 (1999).
[7] D. Ravelosona, A. Cebollada, F. Briones, C. Diaz-Paniagua, M. A. Hidalgo, and F. Batallan, Phys. Rev. B **59**, 4322 (1999).
[8] U. Ebels, A. Radulescu, Y. Henry, L. Piraux, and K. Ounadjela, Phys. Rev. Lett. **84**, 983 (2000).
[9] R. Danneau, P. Warin, J. P. Attané, I. Petej, C. Beigné, C. Fermon, O. Klein, A. Marty, F. Ott, Y. Samson, and M. Viret, Phys. Rev. Lett. **88**, 157201 (2002).
[10] P. M. Levy and S. Zhang, Phys. Rev. Lett. **79**, 5110 (1997).
[11] G. Tatara and H. Fukuyama, Phys. Rev. Lett. **78**, 3773 (1997).
[12] R. P. van Gorkom, A. Brataas, and G. E. W. Bauer, Phys. Rev. Lett. **83**, 4401 (1999).
[13] N. Agraït, A. Levy Yeyati, and J.M. van Ruitenbeek, Physics Reports **377**, 81 (2003).
[14] R. Landauer, IBM J. Res. Dev. **32**, 306 (1988).
[15] L. I. Glazman, G. B. Lesovik, D. E. Khmelnitskii, and R. I. Shekhter, JETP Lett. **48**, 238 (1988).
[16] T. Ono, Y. Ooka, and H. Miyajima, Appl. Phys. Lett. **75**, 1622 (1999).
[17] F. Elhoussine, S. Mátéfi-Tempfli, A. Encinas, and L. Piraux, Appl. Phys. Lett. **81**, 1681 (2002).
[18] F. Komori and K. Nakatsuji, Mat. Sci. Eng. B **84**, 102 (2001).
[19] C.-S. Yang, J. Thiltges, B. Doudin, and M. Johnson, J. Phys.: Cond. Matter **14**, L765 (2002).
[20] J. Velev, R. F. Sabirianov, S. S. Jaswal and E. Y. Tsymbal, Phys. Rev. Lett. **94**, 127203 (2005).
[21] N. García, M. Muñoz, and Y.-W. Zhao, Phys. Rev. Lett. **82**, 2923 (1999).
[22] W. F. Egelhoff *et al.*, J. Appl. Phys. **95**, 7554 (2004); J. J. Mallett *et al.*, Phys. Rev. **70**, 172406 (2004).
[23] H. Imamura, N. Kobayashi, S. Takahashi, and S. Maekawa, Phys. Rev. Lett. **84**, 1003 (2000).
[24] L. R. Tagirov, B. P. Vodopyanov, and K. B. Efetov, Phys. Rev. B **65**, 214419 (2002).
[25] V. K. Dugaev, J. Berakdar, and J. Barnas, Phys. Rev. B **68**, 104434 (2003).
[26] M. Ye. Zhuravlev, E. Y. Tsymbal, S. S. Jaswal, A. V. Vedyayev, and B. Dieny, Appl. Phys. Lett. **83**, 3534 (2003).
[27] J. B. A. N. van Hoof, K. M. Schep, A. Brataas, G. E. W. Bauer, and P. J. Kelly, Phys. Rev. B **59**, 138 (1999).
[28] J. Kudrnovsky, V. Drchal, C. Blaas, P. Weinberger, I. Turek, and P. Bruno, Phys. Rev. B **62**, 15 084 (2000).
[29] B. Yu. Yavorsky, I. Mertig, A. Ya. Perlov, A. N. Yaresko, and V. N. Antonov, Phys. Rev. B **66**, 174422 (2002).
[30] J. Velev and W. H. Butler, Phys. Rev. B **69**, 094425 (2004).
[31] A. Bagrets, N. Papanikolaou, and I. Mertig, Phys. Rev. B **70**, 064410 (2004).
[32] A. K. Solanki, R. F. Sabiryanov, E. Y. Tsymbal, and S. S. Jaswal, J. Magn. Magn. Mater. **272–276**, 1730 (2004).
[33] M. Büttiker, Phys. Rev. Lett. **57**, 1761 (1986).
[34] S. Datta, *Electronic Transport in Mesoscopic Systems* (Cambridge, University Press, Cambridge, UK, 1999).
[35] P. Gambardella *et al.*, Nature (London) **416**, 301 (2002).
[36] J. Hong and R. Q. Wu , Phys. Rev. B **67**, 020406(R) (2003).
[37] N. D. Lang and W. Kohn, Phys. Rev. B **1**, 4555 (1970).
[38] G. M. Pastor and K. H. Bonemann, in *Metal Clusters*, edited by W. Ekardt, *Whily* (1999).
[39] O. Genzken and M. Brack, Phys. Rev. Lett. **67**, 3286 (1991).
[40] A. Smogunov, A. Dal Corso, and E. Tosatti, Phys. Rev. B **70**, 045417 (2004).
[41] K. M. Schep, P. J. Kelly, and G. E. W. Bauer Phys. Rev. B **57**, 8907 (1998).
[42] E. Y. Tsymbal and D. G. Pettifor, Phys. Rev. B **54**, 15314 (1996).
[43] E. Y. Tsymbal and D. G. Pettifor, *Perspectives of giant magneto-resistance*, in Solid State Physics, edited by H. Ehrenreich and F. Spaepen, Vol. **56** (Academic Press, 2001) pp.113-237.
[44] W. E. Picket and J. S. Moodera, Phys. Today **5**, 39 (2001).
[45] See for example, Sears and Zemansky's University Physics, 11th Ed. H. D. Young and R. A. Freedman, p. 1265.9